\documentclass[11pt, a4paper]{article}
\usepackage{amsmath,cite,url,natbib}
\usepackage{graphicx}
\usepackage{color}

\usepackage{geometry}
\bibfont{\scriptsize}

\title{The Rarity of Musical Audio Signals Within the Space of Possible Audio Generation}
\author{Nick Collins\footnote{Durham University Music Department}}
\date{}

\begin{document}

\maketitle
\begin{abstract}
A white noise signal can access any possible configuration of values, though statistically over many samples tends to a uniform spectral distribution, and is highly unlikely to produce intelligible sound. But how unlikely? The probability that white noise generates a music-like signal over different durations is analyzed, based on some necessary features observed in real music audio signals such as mostly proximate movement and zero crossing rate. 
Given the mathematical results, the rarity of music as a signal is considered overall. The applicability of this study is not just to show that music has a precious rarity value, but that examination of the size of music relative to the overall size of audio signal space provides information to inform new generations of algorithmic music system (which are now often founded on audio signal generation directly, and may relate to white noise via such machine learning processes as diffusion). Estimated upper bounds on the rarity of music to the size of various physical and musical spaces are compared, to better understand the magnitude of the results (pun intended). Underlying the research are the questions `how much music is still out there?' and `how much music could a machine learning process actually reach?'.  
\end{abstract}
\section{Introduction}\label{sec:introduction}


Prokofiev was aware of the combinatorial explosion in modelling music \citep{prokofiev78}; as a keen amateur chess player he would also have been familiar with the origin story of the chessboard where the inventor requested a geometric series of wheat as his reward, doubling from an initial single grain for each of the 64 squares ($2^{64} - 1$ being much greater than any wheat supplies available in the world, ancient or modern) \citep{macdonell1898art}. 

A more recent attempt to copyright all possible simple 8 note melodies (naively ignoring rhythm and any microtonality) was outlined by activist lawyer Damien Riehl who claimed to have `exhausted the data set' \citep{Cuthbertson20}. This use of a straw model insufficiently complex to cover the wide variety of musical expressions in the world is a common issue, pace Ed Sheeran's comments following the `Shape of You' copyright court case `There are only so many notes and very few chords used in pop music and coincidences are bound to happen if 60,000 songs are being released a day on Spotify, that is 22m songs a year, and there are only 12 notes that are available.' \citep{Khomami22}. In this article a variety of musical spaces are considered, with respect to all audio signals of a reference 44.1KHz sampling rate and16 bit resolution, rather larger than Sheeran's basic 12 note model, to attempt to enrich understanding of the scope of musical possibility. 

There is a long historical backdrop in mathematical treatments of the combinatorics of music theory \citep{nolan2000musical}. 
It is clear that the mathematical size of music is very large, especially when models are purely mathematical and ignore any culling through perceptual constraints. 
Evidence that within perceptual terms the range of musical options is more limited is given by \citet{frieler2011independent}, in the context of pop melody writing to a given chord sequence by amateurs, when overlaps between participant-composed melodies were discovered. 
From an AI angle, given recent and intensive investigation into sample by sample (neural) synthesis models \citep{oord2016wavenet}, what size of output space is accessible to different algorithms? If an algorithm generates audio directly, rather than operating in a symbolic domain, it may have a substantial advantage in potential width of generation, though in practice its training set, representation and architecture may more severely constrain generation.   

Critical research questions for this study are: 
\begin{enumerate}
\item In what ways can the rarity of musical signals be quantified relative to the space of possible audio signals? 
\item How can the probability of music with respect to a given model of audio be measured? 
\item What proportion of all possible musical audio signals have humans already explored? 
\end{enumerate}
which will only be partially answered in this paper, but are strong motivations for this type of theoretical study.

\section{White noise analysis}\label{sec:whitenoise}

White noise is of interest because one way of sampling from the space of all possible signals is to allow a free selection of the amplitude value at each sample, and this is what the independent draws of such a random noise generator explore. 

What is the probability that a white noise process over $n$ samples will output a music-like signal rather than broadband noise? 
The pathological case where a musician runs a white noise generator for the entire duration of their piece without any refinement is ignored; noise music actually includes quite interesting variation over time anyway \citep{hegarty07}, but the profile of more conventional Western classical and popular music is the baseline. 
The basic digital white noise algorithm calculates independent samples, generated by successive and unconnected draws from a uniform distribution within -1 to 1 or across some other N bit integer range. From running white noise generators in computer music systems the informal understanding is that such generators do not output even short segments of more music-like signal, but are perceptually true to their statistically average full band spectrum. It is theoretically possible (but vanishingly unlikely) to get back any recording that could ever be conceived, of any human (or indeed alien) speaker or musician, any ensemble of any instruments known or unknown, recorded in any conceivable acoustic. 

In order to analyse how likely a white noise process is to coincidentally generate a music signal, two necessary conditions for music signals are examined. 

\begin{itemize}
\item White noise has a 50\% chance of a zero crossing with every new sample. Zero crossing rates for musical signals are much lower. 
\item Musical signals move to close by sample values much more often (are more continuous) than white noise, where `close by' can be quantified. Proximate values within $\epsilon$ of x[n] occur with probability $\epsilon$ for white noise.\footnote{e.g., Moving within an absolute difference $|x[n]-x[n-1]| < \epsilon$} 
\end{itemize}

Through audio analysis of a corpus of music recordings, the observed zero crossing rate was around 5\% rather than the 50\% for basic white noise, and successive float values within 0.1 99.7\% of the time. This is not to say that white noise couldn't (over a short term) generate less zero crossings and more continuous signal, but how likely is this? 

The binomial distribution provides the answer. A 50\% chance of zero crossing is like a fair coin being tossed on independent trials; a 99.7\% chance of proximate movement a rather more biased coin. Given a Bernoulli process over N samples the probability of M `heads' (whether representing zero crossings, or proximate movement) can be calculated, given probability $p$ per trial. 

The probability of exactly $k$ `heads/successes' over $n$ samples, given a probability of heads of $p$ is: 

\begin{equation}
P(k; p, n) = \binom{n}{k} p^k (1-p)^{n-k}
\end{equation}

A sum is taken over all $k \geq K$ where large $K$ is chosen to reflect the minimum number of expected proximate movements (or conversely all $k \leq K$ where large $K$ is chosen to reflect the maximum number of zero crossings). 

\begin{equation}
P(k \geq K; p, n) = \sum_{k=K}^{n} \binom{n}{k} p^k (1-p)^{n-k}
\end{equation}

In order to calculate this, since it involves many very small numbers due to the large exponents for the powers, and very large numbers due to larger binomial coefficients, python's mpmath library (multi-precision math) \citep{mpmath} was used. The chance of white noise generating a one second signal at 44100 sampling rate in the range [-1,1], with 99.4\% or more proximate movement within 0.1 of the previous value (taking $K = 0.994*44100 = 43835$), has a probability of 1.24355865e-2018, that is, a probability vastly smaller than the chance of selecting a single atom in the observable universe (1 in $10^{80}$). 

\begin{figure}[h]
 \centerline{\framebox{
 \includegraphics[width=0.9\columnwidth]{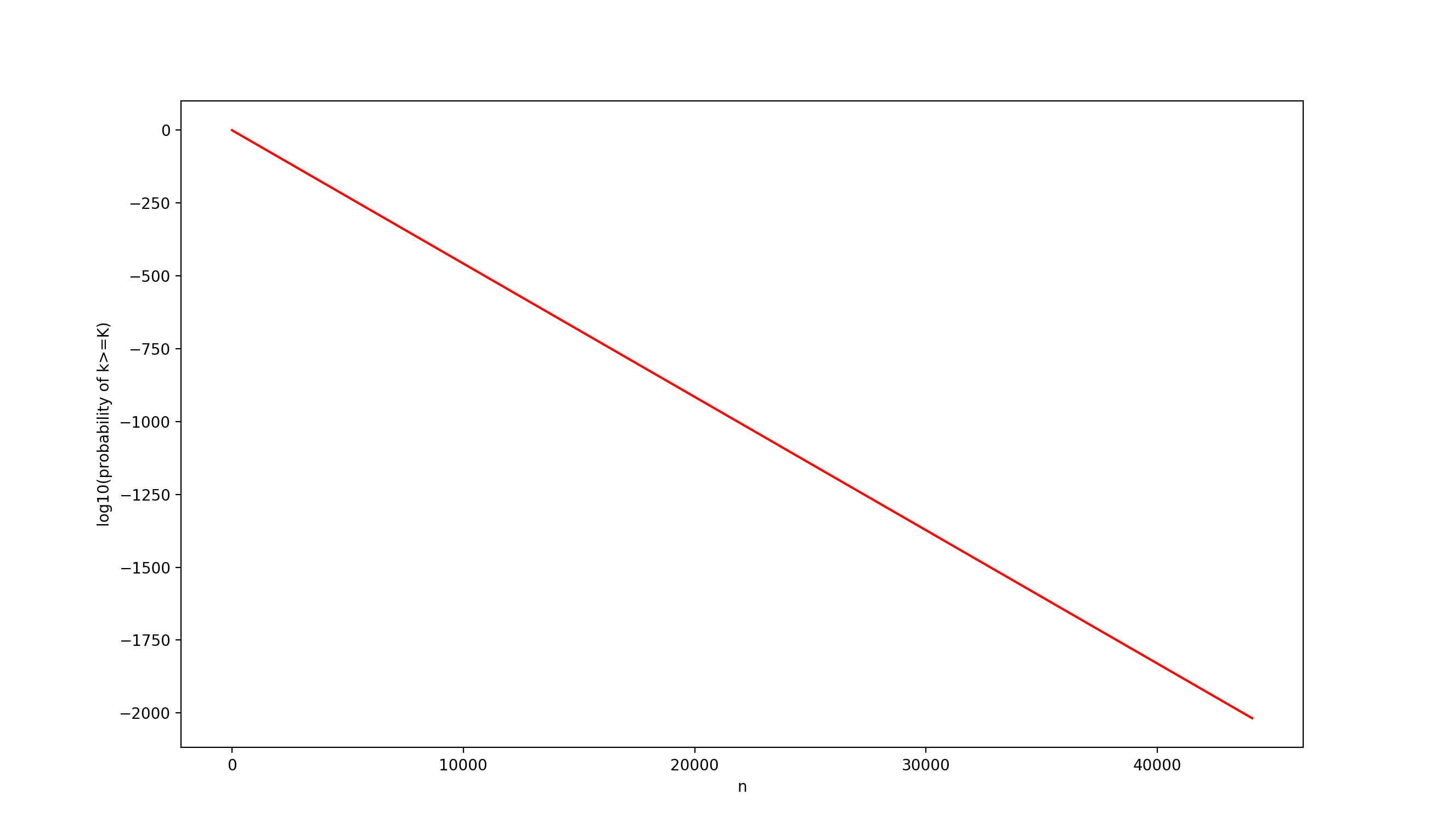}}}
 \caption{Changing log (to base 10) probability as $n$ increases from 2 to 44100}
 \label{fig1}
\end{figure}
\begin{figure}[h]
 \centerline{\framebox{
 \includegraphics[width=0.9\columnwidth]{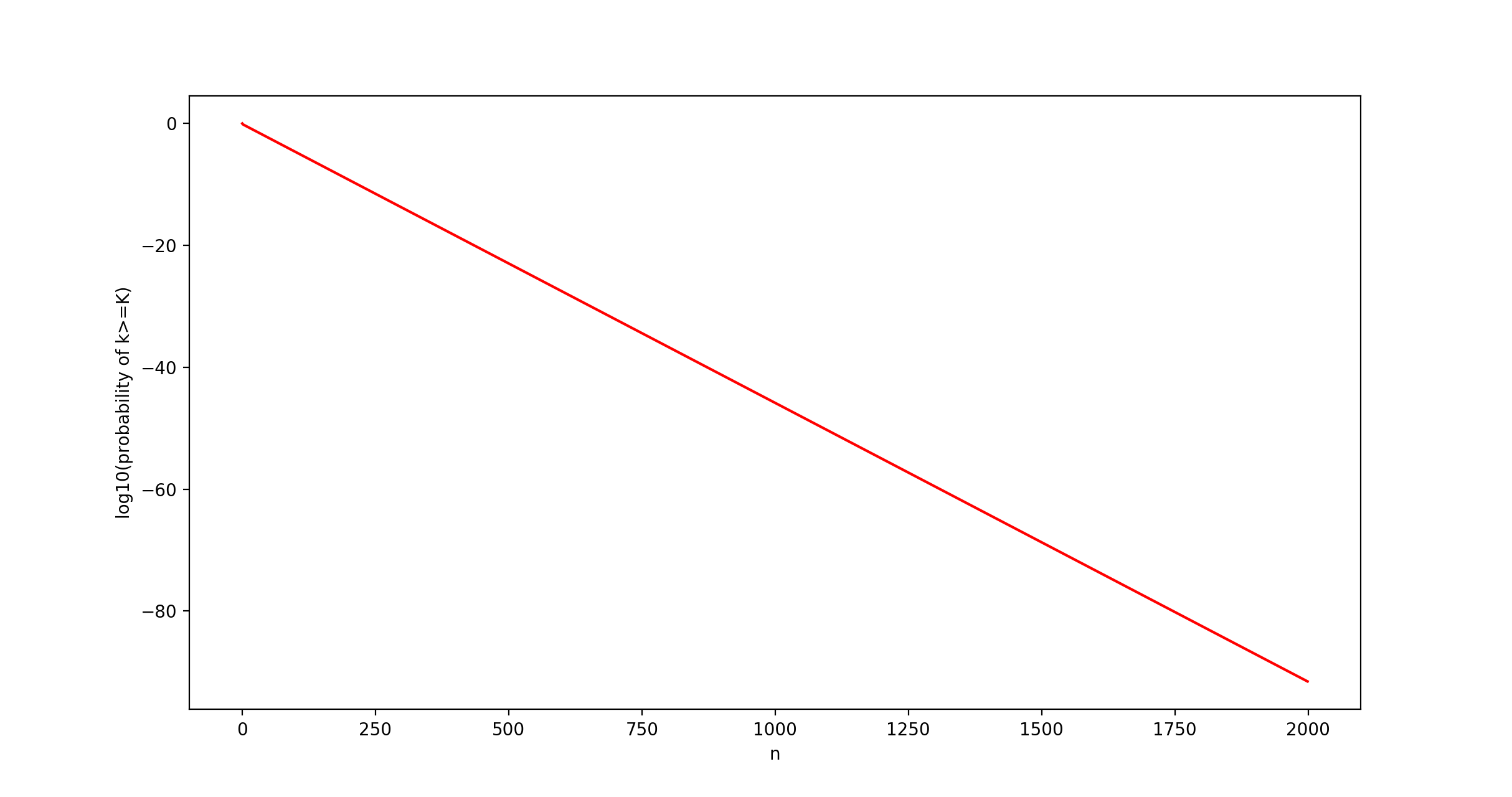}}}
 \caption{Changing log (to base 10) probability as $n$ increases from 2 to 2000}
 \label{fig2}
\end{figure}

This was for fixed $n$ corresponding to one second of audio at 44100 Hz sampling rate. In order to observe how this changes as the $n$ increases, reflecting longer and longer windows of samples that must be minimally music like, figure \ref{fig1} plots the calculation on a $log_{10}$ probability scale for increasing $n$ from 2 to 44100 ($K$ and $p$ being fixed). Zooming in on the range up to $n=2000$ in figure \ref{fig2} shows that the 1 in $10^{80}$ probability is crossed around 1750 samples, that is 39 milliseconds at the 44100 sampling rate (and less time at higher rates). These graphs plotting log probability against n look linear, but the differences from $n$ to $n+1$ vary very slightly (at thousands of decimal places in; this can be confirmed by running the code with mp.dps = 2500, for instance). The near linear increase with $n$ is due to the exceptionally small size of the probability weight for $k \geq K$ relative to the bulk of the weight for other $k$.   

The case for zero crossings is similar; white noise must now generate far less than its usual 50\% zero crossings to match the typical lower zero crossing rate for music. This could be calculated as above, but an alternative is to use the Chernoff bound, an inequality used in analysis of the binomial distribution. The probability of generating no more than $k = 2205$ zero crossings (0.05\% of 44100) is bounded above by 

\begin{equation}
\exp((-n) * (a * \log(a/p)) + ((1-a)* \log((1-a)/(1-p))))
\end{equation}

where $a = k/n$. The upper bound for $n = 44100$ $p = 0.5$ and $k = 2205$ is 4.1484712e-9474.


Alternative statistics could be investigated that are more fully representative of musical signals (for instance, music signals tend to have spectra falling in power with frequency 
), but since the probabilities here are \emph{upper} bounds on the probability (necessary conditions) they demonstrate how difficult it is to generate more complicated music from a white noise generator, and by extension, how small a proportion of all possible signals musical signals typically are.

%
%

\section{How much (recorded) music is there, or could there be, compared to all possible signals?}

\begin{table}[t]
\begin{center}
{\footnotesize
\begin{tabular}{|p{12cm}|p{1.5cm}|p{1.5cm}|} 
\hline
Scenario & Size of Space as power of 2 (to 2 d.p.) & Order of magnitude (rounded off log10(size))\\
\hline 
\hline 
All possible one second 64 bit 192000SR audio segments & 12288000 & 3699057\\
\hline 

All possible one second 16 bit 44100SR audio segments & 705600 & 212407\\
\hline 

Estimated possible input contexts for ChatGPT 4 based on a dictionary of 170000 words and a context of up to 32768 tokens ($170001^{32768}$) & 569350.02 & 171391\\
\hline 

Upper bound on more music-like 16 bit 44100SR audio segments based on signal continuity& 205703.43 & 61923 \\
\hline 

Upper bound on more music-like 16 bit 44100SR audio segments based on zero crossings& 31469.89 & 9473\\
\hline

The number of possible outputs if a reasonable sized orchestra of 30 instruments play one 4/4 measure of music at 240bpm, each with a free choice from a 3 octave range in 12TET or a rest (37 options), and up to 32 different time positions to play at ($37^{32*30}$) & 5001.08 & 1505 \\
\hline

Number of atoms in the observable universe & 265.75 & $ 80$ \\
\hline 

Number of orbit representatives for the (rhythm-pitch) motives in Z12 x Z12 \citep{fripertinger1999enumeration} & 124.66 & 38\\
\hline 	

Number of one second audio fragments created if 1024 microphones surround each of the 100 billion humans who ever lived \citep{haub11}, and record their whole lives with one second snapshots every sample (taking a modern lifespan average of 80 years rather than more brutish historical life spans, running at 44.1KHz) & 98 & 30\\
\hline 

Number of possible monophonic melodies within a two octave span of 12TET pitches (plus one rest token), over 16 steps ($25^{16}$) & 74.30 & 22\\  
\hline 

The age of the universe in seconds (13.787 billion years * 365.25*24*60*60), or the number of beats at 60bpm since the dawn of time&  58.59 & 17 \\

\hline

The number of one second audio fragments created at step size of one sample (at 44.1KHz), if 17 billion mobiles in the world each record constantly for 3 years before they break (and ignoring the practicality of charging downtime etc). & 47.42 & 14 \\
\hline

All 53875981680676 variations accessible to Diederich Winkel’s mechanical Componium premiered in 1821 \citep{Bumgardner13} & 45.61 & 13\\
\hline

SoundCloud catalog size of 200 million tracks, averaging around 3 mins each; all one second audio windows (stepping every sample at 44.1KHz) & 39.74 & 12\\
\hline 


All 9985920 tone rows in a 12 pitch class system \citep{fripertinger1999enumeration} & 23.25 & 7\\
\hline  

All 65536 16-step basic rhythmic patterns (on or off at each step) & 16 & 5\\
\hline 

Around a 1 in 29842 chance of finding a needle when grabbing hay from an average sized haystack (hand grab sphere radius 5cm, haystack side 2.5 metres \citet{haystack}) & 14.87 & 5 \\
\hline


\hline 
\end{tabular}
}
\end{center}
\caption{\small{Comparative sizes of musical resources according to different scenarios}}
\label{musicsizes}
\end{table}%

Table \ref{musicsizes} compares the size of various collections of musical audio, various musimathematical spaces, and various physical systems, with respect to the possibilities for one second of monophonic digital audio. The top entry, for a very high sample rate and bit resolution, far exceeds human perceptual capabilities, and a mixdown to 44.1KHz 16 bit audio (as for example in preparing a CD) would maintain essential auditory properties.\footnote{it could be argued further that a sufficiently high bit rate audio codec, for example a 256kbps MP3, at a size reduction of around 1 in 5, might also do, but this wouldn't save much in order of magnitude}. Considering an N audio channel system would multiply the size, and a lower bit resolution and sampling rate (as in earlier video game music) reduce it, but is not investigated further here. 

It fits intuition that the space of musical signals within white noise is tiny. You'd wait many times longer than the age of the universe to roll the dice once per second to get anything musical as output: ((1.4e10)*365.25*24*60*60) = 4.418064e+17 seconds so far within 14 billion years. There are about 1e81 atoms available, so you can't run lots of parallel computers to get there either.

The combinatoriality of music can be extreme; only 10 options per event over a piece of 80 independently selected events are needed to match the size of the observable universe.\footnote{The number of configurations of those atoms in physical space is vastly larger than anything in the table, however, given a 93 billion light years diameter for possible locations.} This however assumes an entirely free and independent selection, avoiding any particular cognitive preferences such as a tendency to refer more often to a pitch centre, stepwise motion above larger leaps, and other perceptual constraints \citep{lerdahl88, brown15}. 

It is possible to envisage creating larger musical spaces than those outlined here. Consider an N virtual instrument track digital audio workstation project, each synth with M 14-bit parameters (across many sound synthesis and audio effects settings) updated every grid point; there are X possible grid positions per second for rhythmic placement of events. Each event can take on a new pitch from a dense microtonal scale of Y distinct pitches; all synths are on all the time (there are no rests, so no break from automation data) but there is an option to continue the current pitch without a new onset (so Y+1 options for event pitches at each step). Then the size of the musical space of second long fragments within this model is $((Y+1)*16384*M * N)^X$
and comparing the $\log_{10}$ size to all possible one second 16 bit 441.KHz audio segments (of order 212407): 

\begin{equation}
X * ( \log_{10}(Y+1) + \log_{10}(16384) + \log_{10}(M) + \log_{10}(N))
\end{equation}

Choosing the values X = 96 N = 100 M = 100 Y = 99: 

\begin{equation}
96 * ( \log_{10}(100) + \log_{10}(16384) + \log_{10}(100) + \log_{10}(100) ) = 980.58431417239  \ll 212407
\end{equation}

It proves rather difficult to put enough tracks and parameters into such a system to beat audio rates, without setting the rhythmic grid to inhumanly fast sampling rates. However, one possible case that does at least approach greater size is the notion of a very large crowd (perhaps a huge festival crowd baying along to their favourite headliner), with a complicated vocal model, modelling microtonal pitch fluctuation and microtiming X = 2048 N = 1000000 M = 100 Y = 249 

\begin{equation}
2048 * ( \log_{10}(1000000) + \log_{10}(16384) + \log_{10}(100) + \log_{10}(250) ) = 31974.113173438  < 212407
\end{equation}

In order to exceed the one second audio space size, a musical piece might be considered as a whole across many minutes, pushing up the time steps sufficiently to exceed an audio sampling rate over one second; however, any recorded rendering of the piece could be split into one second segments, or a model based on many minutes of audio samples.   
s
So the rarity of music is very high as measured by some notion of `extant recordings’ or even `all possible human lifetimes of recording’.

%
%
%

\section{Conclusions}

Consideration of the ‘The Rarity of Music’ is a productive task in musical modelling, to increase our sense of the preciousness of human musical endeavour. According to the analysis of white noise presented here, musical signals are a very small subset of the space of possible signals. An approximate upper bound is presented; real music would have more conditions than sample to sample proximity or zero crossings, so if anything most non-pathological musics are even rarer compared to white noise. Exactly how rare is further quantified by Table \ref{musicsizes}, but future investigation could further plot this space of possible musics, with a particular ear to models of music cognition. A next stage would be to further quantify the size of perceptual music spaces, that is, the size of musical possibility with respect to various psychological models. 

This work is within a wider context of thinking about how much music there is out there for AIs to populate. In a situation where AI music generation commonly outputs billions of pieces, such that the virtual music world vastly outnumbers the human manually created one, a worry arises that there will be no room for humans to innovate. In actual fact, computers running with all available physical resources in the universe are unlikely to dent the true potential of musical space (as long as overly basic limited pitch set and highly time quantised theories are avoided).

\section{Ethics statement}\label{sec:ethics}

No potential conflicts of interest are anticipated; this research has not involved human participants in any way and a university ethics committee was not required to pre-approve it. 

Nonetheless, there are potential societal impacts to this work, in the context of the wild growth of AI music systems, and increasing access to music composition as a pastime for large numbers of human creators. A central contention of this work is that within sufficiently extended music spaces, the flexibility to create diverse works can outnumber the atoms in the observable universe, and therefore the available computational resources that could ever be thrown at the problem of computer music generation. More work remains to be done to refine estimates of the cardinality of perceptual possibilities rather the purely mathematical ones, however, and further calculations as to the generative scope of music AI in constrained stylistic areas would be necessary to settle certain fears of AI music over-running human musical creativity. Legislative solutions to protect human creators are not investigated here, but would be better informed by study of the `size of music'.

\section{Acknowledgments}\label{sec:ack}

With many thanks to Bob L. Sturm for some initial discussions on this topic. Any errors in this paper are mine alone.

\bibliography{rarity}

\end{document}